\documentclass[sigconf]{acmart}

\setcopyright{none}

\begin{document}
\title{Distributed Symmetric Key Establishment: \break A scalable, quantum-proof key distribution system}

\newcommand{\tsc}[1]{\textsuperscript{\ensuremath{#1}}} 
\newcommand{\tsf}[1]{\textsuperscript{\ensuremath{#1}}} 
\author{Hoi-Kwong Lo\tsc{a}, Mattia Montagna\tsc{a,*}, Manfred von Willich\tsc{a}}
\affiliation{
  \institution{\vskip .2cm} 
  \institution{\tsf{a} Quantum Bridge Technologies Inc., Toronto, Canada}
  \institution{\tsf{*} mattia.montagna@qubridge.io}
}

\begin{abstract}
We propose and implement a protocol for a scalable, cost-effective, information-theoretically secure key distribution and management system. The system, called Distributed Symmetric Key Establishment (DSKE)\footnote{Originally named Quantum Key Infrastructure (QKI) and then Distributed Symmetric Key Exchange (DSKE).  The term `key establishment' is to align with NIST terminology.}, relies on pre-shared random numbers between DSKE clients and a group of Security Hubs. Any group of DSKE clients can use the DSKE protocol to distill from the pre-shared numbers a secret key. The clients are protected from Security Hub compromise via a secret sharing scheme that allows the creation of the final key without the need to trust individual Security Hubs. Precisely, if the number of compromised Security Hubs does not exceed a certain threshold, confidentiality is guaranteed to DSKE clients and, at the same time, robustness against denial-of-service (DoS) attacks. The DSKE system can be used for quantum-secure communication, can be easily integrated into existing network infrastructures, and can support arbitrary groups of communication parties that have access to a key. We discuss the high-level protocol, analyze its security, including its robustness against disruption. A proof-of-principle demonstration of secure communication between two distant clients with a DSKE-based VPN using Security Hubs on Amazon Web Server (AWS) nodes thousands of kilometres away from them was performed, demonstrating the feasibility of DSKE-enabled secret sharing one-time-pad encryption with a data rate above 50\nobreakspace Mbit/s and a latency below 70\nobreakspace ms.
\end{abstract}


\maketitle

\section{Introduction} \label{sec:intro}

Symmetric and asymmetric cryptographic schemes employed for digital communication often assume that a hypothetical adversary is computationally constrained. A classic example is the widely used RSA asymmetric scheme, which assumes that factoring a large integer into its two prime factors is computationally infeasible. For an adversary in possession of the factors, the communication between the parties involved would be completely insecure. Shor discovered an efficient quantum algorithm for factoring large integers \cite{shor1999polynomial}. Shor's algorithm, when run on a quantum computer, can break the RSA scheme, Diffie--Hellman key exchange and elliptic curve cryptosystems, and consequently poses a serious threat to Public Key Infrastructure (PKI). Although a large-scale commercial quantum computer is not yet available, a malicious eavesdropper can readily store data being exchanged today\footnote{Such harvesting of data is commonplace. The Venona project is an early example. (https://en.wikipedia.org/wiki/Venona\_project)} for when new breaches of a protocol are developed, or technological advances make existing theoretical exploits practical. The advent of quantum computers will bring this about, imperiling long-term security, and threatening political, social and economic stability. For these reasons, several governments and international organizations are planning a transition to quantum-safe cryptographic solutions, and NIST is currently standardizing post-quantum cryptographic systems \cite{chen2016report}.

Commercially available quantum-safe cryptographic solutions range from Post-Quantum Cryptography (PQC) algorithms \cite{bernstein2017post,nejatollahi2019post} to Quantum Key Distribution (QKD)\cite{xu2020secure,gisin2002quantum}. PQC algorithms are similar in nature to current cryptographic algorithms such as RSA, since they rely on mathematical problems that are believed to be very difficult to solve, even with the use of a quantum computer \cite{bernstein2017post,buchmann2016post}. PQC comes with no guarantee that a classical or quantum computer will not be able to solve the problem underlying their security. In other words, those protocols are at risk from new classical and quantum algorithms being discovered. QKD, on the other hand, is unconditionally secure, even against quantum computers, but is limited in its key rate and communication distance \cite{xu2020secure,lo2014secure,diamanti2016practical}. Furthermore, QKD appliances are still relatively expensive, and they require dedicated optical fibers or line-of-sight free space channel such as ground-to-satellite to work.

As proven by Shannon \cite{shannon1949communication}, one-time-pad (OTP) encryption offers information-theoretical confidentiality. Where two communicating parties have a random key that is (i) known to them only, and (ii) as long as the message to be exchanged, they can safely communicate by using this key to encrypt and decrypt the message, and then erase the key. There is no means by which an eavesdropper listening to their conversation can distinguish the intercepted messages from messages that contain randomly generated data. Other symmetric encryption ciphers, like the Advanced Encryption Standard (AES), do not demand keys with the same length of the message, at the cost of losing the information-theoretical security property. The problem of securely distributing the key to the communicating parties, known as the \textit{key distribution problem}, remains. One could imagine that a service provider may fulfill this service by distributing keys to any pair or group of people who require them for communication using symmetric encryption. Though clearly possible, this solution would require complete trust in the provider, who could violate that trust and access all the communication between the parties. It would also require anticipating every communication link and providing keys for that link to the parties ahead of time, which can carry $O(q^2)$ cost or worse, where $q$ is the number of communicating parties\footnote{For example, if $q = 1000$ individuals, and all potentially have a need to communicate securely in pairs, there would have to be $q(q - 1)/2 = 499500$ keys previously distributed. For arbitrary groups of these needing to communicate securely, this increases to $2^q - q - 1$ keys.}. We note that pre-shared keys are a useful cryptographic solution in some applications, for example Wireless Sensors Networks (WSNs), where devices are resources constrained and asymmetric key cryptosystems are unsuitable \cite{chan2003random}. Furthermore, pre-shared keys are mentioned as a viable quantum-safe cryptographic solution by the NSA \cite{NSA}.

Alternatively, if a trusted third party exists, then one can construct a star network with the trusted third party being the central node and be responsible for sharing keys with each user. Such a design is scalable as each user needs only to share a key with the central node, called Charles. When two users, named Alice and Bob, want to communicate with each other, the central node, Charles, can do \textit{privacy swapping}. By simply taking the exclusive-or of the two keys between Alice and Charles and between Bob and Charles and publicly broadcasting it, Alice and Bob can now share key among themselves. The only drawback is, of course, Charles now has full knowledge of the key. Incidentally, such \textit{privacy swapping} idea is also widely used in the trusted relay network design in standard QKD networks in China and Japan. Therefore, those networks are vulnerable to attacks due to compromised relay nodes\footnote{In the context of QKD, a method to distribute trust was proposed in \cite{curty2019foiling} and implemented in \cite{li2021experimental}.}. In summary, how to build a robust, scalable network with distributed trust that is cryptographic agile and quantum-safe is a challenge in the field.

In this paper, we propose a design based on a \textit{robust secret sharing scheme}\footnote{An $(n,k)$ secret sharing scheme is a method to encode a secret data string $S$ into $n$ shares, such that $S$ may be reconstructed from any $k$ shares, but $k - 1$ shares reveal no information about $S$, where $1 \leq k \leq n$ \cite{shannon1949communication}. A robust secret sharing scheme is a secret sharing scheme in which a bounded number of malicious participants cannot prevent the reconstruction of the secret through withholding or modifying their shares, even if they are in coalition \cite{cheraghchi2019nearly}.} that solves the problem of key distribution and, at the same time, distributes trust among a number of providers \cite{lo2021key}. This system, which we call Distributed Symmetric Key Establishment (DSKE), is a scalable, information-theoretically secure key distribution system that can be used with the current Internet infrastructure to distribute disposable symmetric secret keys among any group of users \cite{cheraghchi2019nearly}. DSKE is provably quantum-secure. The cryptographic keys obtained via DSKE can then be used to protect data at rest or in transit via standard symmetric ciphers and authentication techniques. In case those techniques are information-theoretically secure, the entire communication maintains the same property.

The DSKE system relies on a configurable number of parties, called \textit{Security Hubs}. When a new client joins the DSKE system, each Security Hub physically ships its high-quality random data\footnote{Note that high speed quantum random generators -- in excess of 1\nobreakspace Gbps -- have been successfully implemented \cite{qi2010high,zhang201568}.} on a secure data storage device, called a Pre-Shared Key Module (PSKM), or via QKD links, when these are available. When two or more DSKE clients want to generate a key, Security Hubs support their request by providing suitable instructions to the clients over a public channel. Importantly, only an adversary who is able to simultaneously compromise sufficiently many Security Hubs will be able to determine the final keys delivered by the DSKE system. And only if the adversary can disrupt a sufficiently large number of communication links will he be able to prevent the reconstruction of those keys. In other words, the DSKE system is resilient against faulty Security Hubs and denial-of-service (DoS) attacks. Those threats may be mitigated by assigning the Security Hub role to multiple private and public corporations or parties within the same organization. The DSKE protocol makes possible a key distribution system whose security does not rely on assumptions on the computational limits of an adversary and can be considered safe against any future potential attacks, by any classical or quantum computer.

In contrast to most other key distribution systems, DSKE is a quantum-secure key distribution system, and not merely quantum-safe, even against future attacks on data intercepted and stored today. This is a critical and major difference with respect to PQC solutions. Our system has the guarantee that, when used with information-theoretically secure encryption and authentication, data remains safe during its entire life cycle with only limited assumptions. The DSKE system also protects its clients from vulnerabilities in the implementation of legacy encryption systems. DSKE is based on very simple and well-understood principles, so its coding and auditing can be easily performed by third parties in a cost-effective manner. DSKE finds applications in network security, end-user device security, and embedded systems. Indeed, the entire protocol can be effectively coded to run on computationally constrained devices where asymmetric encryption is too demanding.

In the following, Sec. \ref{sec:dskeprot} introduces the DSKE protocol; Sec. \ref{sec:adapted} describes an adaptation of the protocol for bandwidth reduction; Sec. \ref{sec:secanalysis} presents the security analysis of the DSKE protocol and discusses how to select implementation parameters;  Sec. \ref{sec:implementation} discusses the DSKE protocol implementation and how to integrate the protocol in existing infrastructure; Sec. \ref{sec:conclusion} concludes.

\section{The DSKE Protocol} \label{sec:dskeprot}

We discuss two versions of the same protocol, based on the same principles and ideas, with the second one adapted for efficiency when DSKE clients are allowed to communicate among each other via insecure public channels. We start by presenting the first version, and leave the adapted version for the next section.

We consider two parties, Alice ($A$) and Bob ($B$), who want to use the DSKE system to generate keys, and a group of $n$ Security Hubs, who are individually untrusted entities labeled $P_1, P_2, \dots , P_n$. The DSKE system can be easily extended to any number of clients and, as we will show, new clients can be added to the system at any time. The role of the Security Hubs, in a nutshell, is to (i) generate high-quality random numbers and physically ship those to DSKE clients while keeping a local private copy; (ii) acts on DSKE client requests by providing key instruction messages that allow them to find a mutually agreed key. At the beginning of the protocol, when a new client joins the DSKE system, Security Hubs generate random numbers using a high-quality entropy source, and deliver a copy of these random data to the client using, for example, a secure data storage device, called a Pre-Shared Key Module (PSKM). Later, when a client wants to agree on a key with one or more others, she communicates this to the group of Security Hubs. Each Security Hub in response sends a key instruction message to the other client(s), who then use these to distill keys from the random data that was pre-shared with them by the Security Hubs when they joined the system. After the distribution of pre-shared keys, the subsequent communication between DSKE clients and the Security Hubs can take place over unsecure, public channels. The key generated by DSKE can then be consumed at will by the clients, for example to seed symmetric encryption algorithms or authentication schemes. The DSKE system, as we will show, achieves the following:
\begin{itemize}
 \item \textit{Confidentiality}: the DSKE clients do not have to trust more than a chosen number of the Security Hubs, since the remaining Security Hubs, even in collusion, cannot determine the final keys of the clients.
 \item \textit{Information-theoretic security}: an adversary in possession of unlimited computational resources is not able to break the system.
 \item \textit{Fault-tolerance}: the DSKE system can tolerate a (configurable) number of faulty Security Hubs, either due to their malevolent behavior or simple malfunctioning.
\end{itemize}

The DSKE protocol can be divided into four main phases:

\begin{enumerate}
 \item PSKM generation and distribution
 \item Peer identity establishment
 \item Key agreement
 \begin{enumerate}
  \item Share generation
  \item Share distribution
  \item Key reconstruction
 \end{enumerate}
\item Key validation
\end{enumerate}

We now discuss each phase in detail.

\subsection{PSKM generation and distribution}

This phase serves to establish trusted identities and share high-quality, secret random data between each Security Hub--client pair. When a new client joins the DSKE system, each Security Hub generates a large number of random bits using a high-quality random number generator. Each Security Hub maintains a local copy of the random data, and securely delivers\footnote{Here, we assume that the delivery process has the necessary security without elaborating. Required security properties are minimally that the sender and receiver identities are authenticated, and that breach of security is detected and notified to either party before their use of the random data.} a copy to the client encoded on a PSKM \cite{tehranipoor2011introduction}. The PSKM additionally includes safeguards against reading by the wrong party, reading data twice, and other measures as may be required to guarantee the secure delivery of the data to the client and detection of any unauthorised access. The quantity of data delivered on the PSKM may be terabits per delivery (or arbitrarily large). The client acknowledges safe receipt of the PSKM through a trusted channel before it is used by either party.\footnote{In a realistic implementation of the DSKE protocol, each Security Hub can include one Security Server, and a number of Local Distributors located in different geographical regions. Each Local Distributor shares a copy of some unallocated random data with its Security Server. These copies are usually physically delivered between a Security Server and its Local Distributors. Local Distributors then deliver the unallocated random data to clients, and notify the Security Server of which portion of the shared random has been delivered to which client. This delivery can be via physical shipping, or, in case a QKD channel is available between a Local Distributor and a client, the Local Distributor can also deliver the random data via this channel. The architecture details of the Security Hub are not discussed in this paper in detail, as they are more of a logistic nature rather than touching the core DSKE protocol. It should be noted, however, that QKD finds a natural place in the DSKE protocol.}

This leaves each Security Hub--client pair with uniquely generated shared secrets in the form of a table of random bits. The number of tables that each Security Hub must maintain is at least equal to the number of DSKE clients that it serves. At each client, the number of tables is limited to the number of Security Hubs that provide a service to that client at that point in time. Importantly, every random bit in the table is only used once in the DSKE protocol. When all the random data in the table is consumed, this first phase is repeated, and the Security Hub delivers a new PSKM to the client.

\begin{figure}[ht]
\includegraphics[width=1\columnwidth]{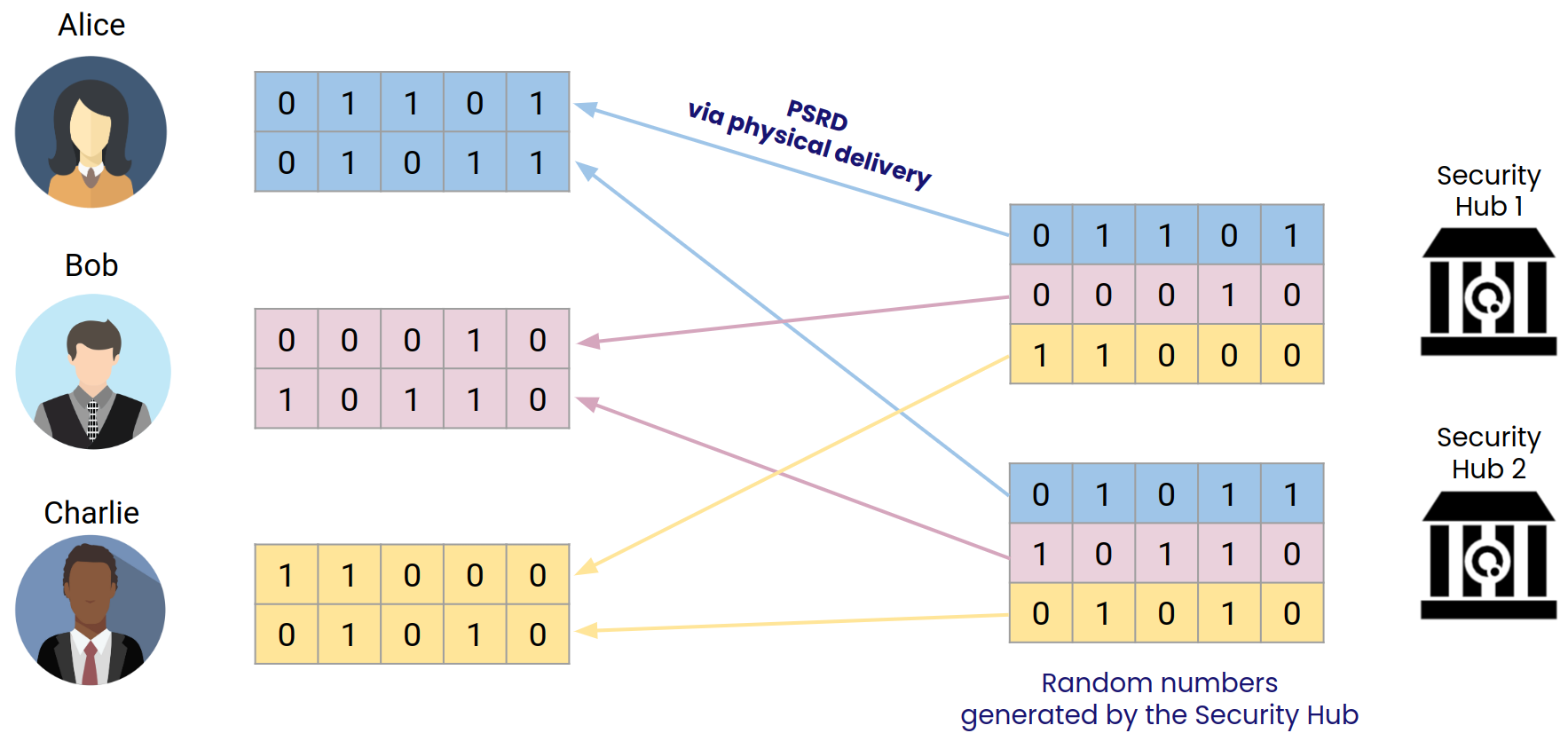}
\caption{\textit{Results of the PSKM generation and physical distribution phase. DSKE clients Alice, Bob and Charlie share an ordered table of bits with each of the two Security Hubs. Each Security Hub only knows his own part of the users' tables.}}
\label{fig:pskm}
\end{figure}

At the end of this phase, $A$ (Alice) shares table $H_1^A$ with $P_1$ (Security Hub 1), $H_2^A$ with $P_2$ (Security Hub 2), and so on. For simplicity, we organize these tables into a matrix $H^A$, with dimensions $n \times N$ (where $n$ is the number of Security Hubs and $N$ is a large number, being the size of each table), and with elements in $\left\{ 0,1 \right\}$. We call the matrix that is shared between $B$ (Bob) and the Security Hubs $H^B$, i.e., the first row $H^B_1$ (of length $N$) is shared between $B$ and $P_1$, the second row $H^B_2$ between $B$ and $P_2$, and so on. Fig. \ref{fig:pskm} provides a representation of the results coming from phase 1 of the protocol. In a nutshell, $A$ and $B$ now each share an ordered table of bits with each of the Security Hubs, and each provider only knows his own part of the client's tables $H^A$ and $H^B$.

Each of the Security Hubs and clients tracks which bits in each table of pre-shared random data that they hold have been consumed (used). At the beginning, when the tables are freshly distributed, all bits are marked as `unused'.

\subsection{Peer identity establishment}

In the initial part of the DSKE protocol $A$ and $B$ need to establish the authenticity of each other's identities. Note that authentication of identities between DSKE clients and the Security Hubs is a result of the PSKM generation and physical distribution phase. Therefore, any DSKE client can use a one- or two-message protocol to query the identities of other clients from each of the Security Hubs, where the Security Hub authenticates and the client validates the identity claim by the Security Hub using an information-theoretically secure message tag\footnote{A message tag is often referred to as a ``message authentication code'', used for data validation.}. A Security Hub that provides identity information about a client that conflicts with a consensus is excluded from the protocol by the querying client. In this way, the initial authentication between clients and Security Hubs can be propagated in the system to authenticate DSKE clients with each other, and each client gets to know the identifier used by each Security Hub for another client. Note that methods to authenticate two parties based on information-theoretic security exist \cite{carter1979universal,alomair2009information}.

\subsection{Key agreement}

This is the core phase of the DSKE protocol, as it allows Alice and Bob to generate a shared key, S. At this stage, each DSKE client shares a certain amount of independent random data with each Security Hub, and all parties have mutually authenticated identities via each Security Hub.

\begin{figure}[ht]
\includegraphics[width=1\columnwidth]{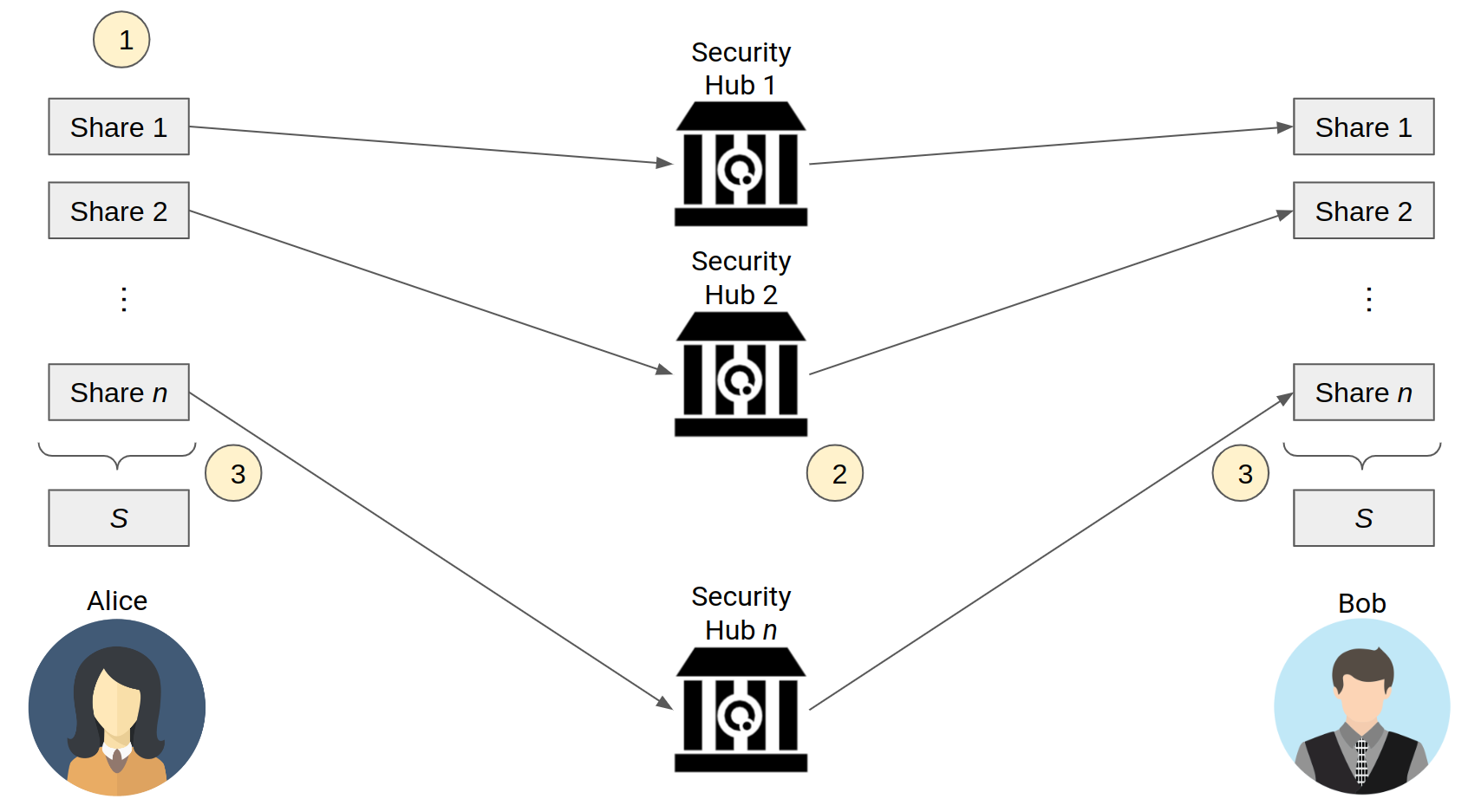}
\caption{\textit{In share generation (1), Alice generates $n$ shares, one associated with each Security Hub, which she will use to build the key using an $(n,k)$ secret sharing scheme. In share distribution (2), Alice transmits each share to Bob using the associated Security Hub, with pre-shared secret random data being used to secure the communication from Alice to the Security Hub, and from the Security Hub to Bob. In key reconstruction (3), both Alice and Bob combine their local shares into a final key $S$ using the same $(n,k)$ secret sharing scheme.}}
\label{fig:comms}
\end{figure}

Suppose that Alice and Bob want to obtain a shared key $S$, of length $m$ bits, via the DSKE system, and that the process is initiated by Alice. Briefly, the key agreement process is the following. First, Alice uses a pre-agreed family of $(n,k)$ secret sharing schemes to generate $n$ shares of the final key $S$. Each share is associated with a specific Security Hub. This phase is called \textit{share generation}. Second, Alice delivers the shares to Bob, using a different Security Hub for each share. The delivery is done leveraging the fact that both Alice and Bob already share secret data with each Security Hub, that can be used to encrypt their communication using OTP encryption. This phase is called \textit{share distribution}. Finally, both Alice and Bob use the pre-agreed family of secret sharing schemes to reconstruct the desired key $S$ from the available shares. This phase is called \textit{key reconstruction}. The entire process is represented in Fig. \ref{fig:comms}.

In the protocol, we will use two main tools, well studied in cryptography: a family of $(n,k)$ secret sharing schemes, and a family of 2-universal hash functions $\{\mathcal{H}_h\}$ \cite{carter1979universal} to generate tags for the messages. A tag $t_{h,M}$ for a message $M$ is a string of $r$ bits generated through one-time use of $h$ of length of $l$ bits, the parameter used to select the function $\mathcal{H}_h$ from the family of hash functions:
$$t_{h,M} = \mathcal{H}_h(M)$$
Such a tag provides information-theoretically secure authentication of the message content. The probability of a single forgery being accepted is at worst $2^{-r}$, the worst-case collision probability of the family of 2-universal hash functions \cite{carter1979universal}.

In the DSKE protocol, bits from the tables of random bits can be marked as `used' by a party as soon as they have read them, though in the case of a failed message validation, this can lead to DoS. To mitigate this, strategies may be developed for marking bits with use counters to permit a limited number of failed validations using the same bits, but will not be addressed in the paper.

We now describe each phase in detail.

\subsubsection{Share generation} \label{subsec:sharegen}

As a first step, Alice needs to decide the length of the key, i.e., the number of bits $m$ in the key that she wants to share with Bob. Alice also chooses a family of $(n,k)$ secret sharing schemes that will be used by her and Bob to rebuild the key. Finally, Alice needs to choose $n$, the number of shares in the scheme, and $k$, the minimum number of shares necessary to reconstruct the key shared with Bob. The choice of $k$, as one may expect, has important implications to the security of the DSKE system, and will be discussed extensively in Sec. \ref{sec:secanalysis}.

At this point Alice is ready to generate the key shares that will be used to distribute the key to Bob. For each $i \in \{ 1, \ldots , n \}$, she selects an unused sequence $R^A_i$ of $m + l$ bits from $H^A_i$, which may simply be the next unused sequence in her table. The integer $l$ is the bit-length of the parameter $h$ for the family of universal hash functions that is used in this protocol for generating a key tag. Therefore Alice arrives at $n$ strings of bits like the following:
$$R^A_i = \left[ \begin{array}{cccc} R^A_{i,0} & R^A_{i,1} & \ldots & R^A_{i,(m+l-1)} \end{array} \right]$$
$$= \left[ \begin{array}{cccc} H^A_{i,j} & H^A_{i,(j+1)} & \ldots & H^A_{i,(j+m+l-1)} \end{array} \right]$$
where $j$ is the index in table $H^A_i$ of the first bit that has never been used before in the DSKE protocol.

Starting from the $n$ strings $R^A_i$, Alice now needs to build $n$ shares, which we call $Y_i$, to send to the Security Hubs. She arbitrarily selects $k$ Security Hubs for which she sets $Y_i = R^A_i$ and uses this as a share for $P_i$.\footnote{Fewer than $k$ shares may be determined in this way, in which case Alice may choose the secret.} In case $k = n$, all the shares have been generated. In the case where $k < n$, she uses these $k$ shares and the secret sharing scheme to construct the remaining $n - k$ shares. At this point, $A$ will have one share $Y_i$ for each $i \in \{1, \ldots, n\}$. Examples of secret sharing schemes are provided in the next section. At this stage, Alice has all the shares ready to build the final secret using the $(n,k)$ secret sharing scheme.

\subsubsection{Share distribution}

The goal of this phase is for Alice to deliver the key shares $Y_i$ to Bob, with the support of the Security Hubs.

Unless Alice will separately communicate a key tag to Bob (as she does in phase 2 of the adapted protocol below), she determines the secret $Y_0$ of $m + l$ bits from the secret sharing scheme, and partitions $Y_0$ into a parameter $u$ of $l$ bits, and a remainder $S$ of $m$ bits. She also  generates a key tag $t_{u,S} = \mathcal{H}_u(S)$, keeping $S$ as the key that is to be agreed.

For each $i \in \{1, \ldots, n\}$, Alice encodes $n$, $k$, the ordered indices of the sequence of bits used from $H^A_i$ to produce $R^A_i$ (which may simply be a start index and a length), along with $B$'s identity, a key identifier (which may include an index into a running key) and the key tag if generated, plus auxiliary information, such as the coordinate $x_i$ for Shamir's secret sharing scheme. She encrypts the $n - k$ shares $Y_i$ that were not generated from $R^A_i$ by using $R^A_i$ as a one-time pad to produce the encrypted share $Z^A_i$ (e.g., $Z^A_i = Y_i \oplus R^A_i$), which she includes in the message. She adds a message tag, which consumes a further $l$ bits of $H^A_i$, thereby bringing the amount used to $m + 2l$ bits, and transmits the aggregate to $P_i$. Thus, $n - k$ of the messages produced by $A$ will be approximately $m + l$ bits longer than those for the other $k$ messages.
Each Security Hub $P_i$ interprets the message that it receives from $A$, checking indices for overlap, validating the message tag and decrypting each share, marking as `used' the indices into $H^A_i$ and validates the message tag. Finally, $P_i$ marks all the bits of $H^A_i$ consumed in the process as `used'.

If the validation of the message from Alice is successful, each Security Hub then determines its share $Y_i$ (in the example, $Y_i = R^A_i$ or $Y_i = Z^A_i \oplus R^A_i$ as determined by the message from $A$). Much like $A$ did, $P_i$ selects an unused sequence $R^B_i$ of $m + l$ bits from $H^B_i$, and uses $R^B_i$ as a one-time key to encrypt the share $Y_i$ to produce $Z^B_i$ (e.g., $Z^B_i = Y_i \oplus R^B_i$). $P_i$ then builds a key instruction message for $B$ as $A$ did for $P_i$, including an encoding of the sequence of indices of the sequence of bits used from $H^B_i$, $A$'s identifier, the key identifier, the encrypted share $Z^B_i$, the key tag that $A$ may have included and a message tag, which consumes another $l$ bits of $H^B_i$. $P_i$ then sends this message to $B$.

For each key instruction message received by $B$ from $P_i$, $B$ checks that the $P_i$ are members of an acceptable set (e.g., a pre-agreed set of $n$ Security Hubs, or else Security Hubs known to serve both $A$ and $B$). $B$ performs the same sequence of actions as $P_i$ did, namely checking that $A$ is an accepted identifier and that $n$ and $k$ (as encoded in the message) are each in an acceptable range (in particular that $k_B \le k \le n$, for a pre-set lower bound $k_B$), checking indices for overlap, validating the message tag and decrypting each share, marking as used the indices into $H^B_i$. Messages that fail in any respect are discarded. At the end of this phase, $B$ has successfully received a number of shares, $s$.

\subsubsection{Key reconstruction}

At this stage Bob has enough information to either reconstruct a set of candidates for the secret $Y_0$, or abort the protocol. When the number of shares $s$ that Bob has received is less than $k$, Bob aborts the protocol. Otherwise, he reconstructs a candidate secret from each subset of $k$ shares with consistent protocol parameters (including the key tag, if present) using the associated $(n,k)$ secret sharing scheme\footnote{There may be up to $^nC_k$ ($n$ choose $k$) such sets, potentially rendering this inefficient as described. Unless there is an active attack by a Security Hub or $A$, one candidate reconstruction followed by checking the other successfully received shares is equivalent and is efficient.}. The next section describes the final phase, which allows Bob to pick a candidate $S$ that can be used for communication with Alice.

\subsection{Key validation}

The goal of this phase is for Bob to either arrive at a final key $S$ shared with Alice, or, alternatively, abort the protocol. 

To begin, $B$ matches all reconstructed candidate secrets against the key tag applicable to the subset of $k$ shares that it was derived from, eliminating those that mismatch. If the number of compromised Security Hubs does not exceed the injection threshold (defined and discussed below in Sec. \ref{sec:secanalysis}), no candidate secret will be known to an adversary, who will consequently be unable to produce a key tag that $B$ will accept, and only one candidate key will remain, albeit from multiple combinations.

More than one distinct candidate remaining implies that the injection threshold has been exceeded, and $B$ aborts the protocol. If $B$ is left with one candidate, this must be $S$, and Alice and Bob correctly conclude the protocol with a shared key, as intended.

\subsection{Specific secret sharing schemes}

In the previous section, we assumed that Alice and Bob agreed on a family of $(n,k)$ secret sharing schemes to rebuild the key $S$ from the shares $Y_i$. We now discuss two concrete examples of those schemes. The first is a general scheme that allows for any $(n,k)$ combination, where $1 \le k \le n$. We call the DSKE protocol using this scheme the \textit{General DSKE Protocol}. In the specific case where $n = k$, there is a significant simplification possible in the implementation of the protocol, and we call this version the \textit{Simple DSKE Protocol}.

\subsubsection{General DSKE Protocol}

Shamir's secret sharing scheme \cite{shamir1979share} is a suitable $(n,k)$ scheme for use in this protocol. For binary data, the field $\mathrm{GF}(2^w)$ is suggested, where $2^w > n$. In most instances, $w = 8$ would suffice, and a field element would correspond to 8 bits. $A$ assigns a unique nonzero value $x_i$ in the field to each $P_i$, $i \in \{1, \ldots, n\}$ and $x_0 = 0$ corresponding to the secret $Y_0$. $A$ divides $P_i$'s share into a sequence of $w$-bit values $y_{i,p}$, where $p \in {0, \ldots, m-1}$ is the offset into the share $Y_i$. Using the same $x_i$ throughout, Shamir's secret sharing scheme is then applied to each offset $p$.

\begin{figure}[ht]
\includegraphics[width=1\columnwidth]{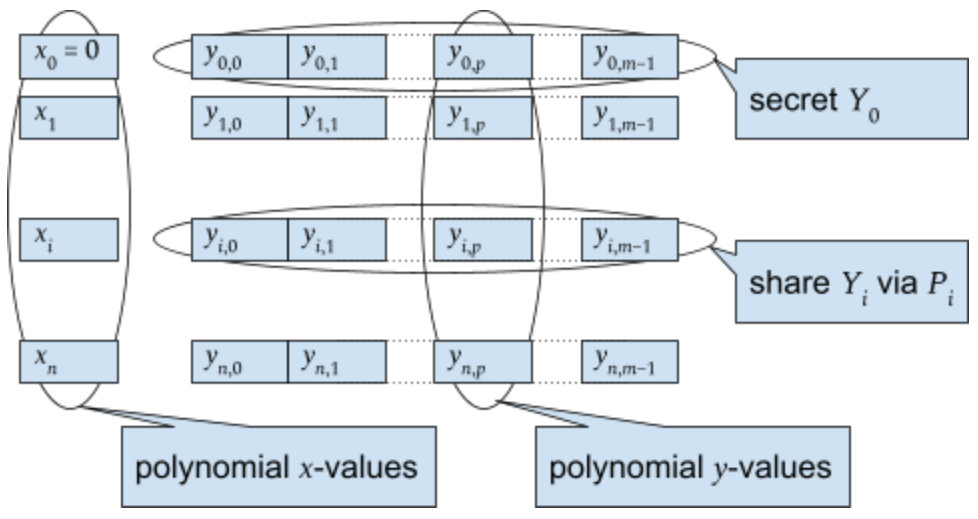}
\caption{\textit{Share segmentation for Shamir's secret sharing scheme.}}
\label{fig:shamir}
\end{figure}

In this scheme, we have vectors $Y_p = \left[ \begin{array}{ccc} y_{0,p} & \cdots & y_{n,p} \end{array} \right]$ and $A_p = \left[ \begin{array}{ccc} a_{0,p} & \cdots & a_{k-1,p} \end{array} \right]$. We also form the matrix
$$X = \left[ \begin{array}{cccc}
              1 & 1 & \cdots & 1 \\
              x_0 & x_1 & \cdots & x_n \\
              & & \vdots & \\
              x_0^{k-1} & x_1^{k-1} & \cdots & x_n^{k-1}
             \end{array} \right]$$
with the matrix equation $Y_p = A_p X$. Note that $A_p$ has dimension $k$, whereas $Y_p$ has dimension $n + 1$. This allows solving for $A_p$ by trimming $Y_p$ to $Y_p'$, including only $k$ known components and keeping only the corresponding columns of $X$, to produce a $k \times k$ matrix $X'$: $A_p = Y_p' X'^{-1}$.\footnote{The matrix $X'$ is always invertible when the values $x_i$ are distinct for distinct $i$ \cite{lai2004several}.} Lagrange polynomial interpolation is a practical method of determining the secret and other shares without the intermediate steps of inverting $X'$ or solving for $A_p$.

\subsubsection{Simple DSKE Protocol}

In this case, $n = k$, and a simple choice for the $(n,n)$ secret sharing scheme is the bitwise exclusive-or operator. The shares generated in the Share Generation Phase (section \ref{subsec:sharegen}) of the DSKE protocol are simply the random bits read from the tables shared with the Security Hubs: $Y_i = R^A_i$, and thus in the share distribution phase, it is sufficient for Alice to send each Security Hub the index $j$ and the length $m$ used to build the share $Y^A_i$. This substantially reduces the amount of data communicated by Alice to the Security Hubs.

\section{Adapted protocol} \label{sec:adapted}

We now present a version of the DSKE protocol that is adapted for improved bandwidth efficiency in the case that Alice and Bob are able to communicate with each other over a public channel. The channel between Alice and Bob is not required to have properties of confidentiality or integrity. The purpose of the adaptation is to minimize the amount of data transmitted by $A$ during the protocol when a very large key is to be agreed, but without compromising the security properties of the general DSKE protocol, other than a potential denial-of-service through interference with the additional bidirectional link. In other words, the Adapted DSKE protocol achieves the robustness against disruption of the General DSKE Protocol with $k < n$, combined with the small amount of data transmitted by $A$ during the protocol when $k = n$, provided that the link used for share negotiation is not disrupted, or short, the efficiency of the Simple DSKE Protocol, with the resilience of the General DSKE Protocol.

The idea of the adapted protocol is the following. First, Alice and Bob use the General DSKE Protocol to agree on a certain amount of key material. Second, Alice and Bob use a simple protocol, but post-selecting which of the shares to use in the reconstruction. Indeed, without the symmetric keys shared between Alice and Bob, the only way to achieve robustness is to use an $(n,k)$ secret sharing scheme with $k < n$, which increases the communication bandwidth from Alice to the Security Hubs. On the other hand, when shares can be securely agreed upon bilaterally, Alice and Bob can then use a simple secret sharing scheme.

In more detail, the General DSKE Protocol is used in a first pass with $A$ choosing values of $n$, $k$, and which Security Hubs to use. In this first use of the DSKE General Protocol, Alice and Bob  generate enough initial key material ($m' = l \times (n + 2)$ bits) for use in the subsequent negotiation of shares for the Simple DSKE protocol, as detailed in the following. Once Alice and Bob established the initial key material, they can use it to securely communicate with each other the shares to use in future calls of the Simple DSKE Protocol. In particular, new shares of size $m + l$ bits (where $m$ is the desired number of resulting agreed key bits) are distributed by $A$ as for the Simple DSKE protocol, and using the same Security Hubs as for the first pass, without a key tag, which will be communicated later during the negotiation. $B$ validates these, additionally rejecting those that do not come from the Security Hubs of the first pass.

Once $B$ has collected enough validated shares to complete the protocol\footnote{This is at least the maximum of $k$ and $k_B$ shares, where $k$ is the value referred to above and $k_B$ is the lower bound on $k$ that is enforced by $B$, though more may be necessary for the protocol to succeed under conditions of attack}, $B$ generates a tag for each of these shares using a separate portion of the initial key so that Alice can verify that these shares were received correctly\footnote{Each share's tag needs key material that is unknown without at least $k$ Security Hubs colluding; otherwise a compromised Security Hub would be in a position to substitute a share that produces the same tag.}. He builds a message that consists of an identifier (such as a Security Hub's $x$-coordinate from the first protocol) with each share's tag, and attaches a message tag, for which he again uses some of the initial key. Thus, $B$ potentially needs $l \times (n + 1)$ bits of key. He sends the message to $A$, who validates each tag for the corresponding share, and draws up a list of $n'$ Security Hubs for which the tags are valid.

If $n' \ge k$, $A$ reconstructs a secret from the shares for this list using the simple $(n',n')$ secret sharing scheme reconstruction. $A$ produces a key tag for this secret, adds it to the list, adds a message tag and sends it to $B$. $A$ needs $l$ bits of the initial key for the message tag. $B$ validates the message tag, checks that the list is a subset of his, reconstructs the secret using the list, and validates the resulting key tag. If $B$ succeeds, $A$ and $B$ have agreed a key of $m$ bits, and consumed up to $m'$ bits of the key that was agreed initially.

Note that $B$ cannot use the Security Hub share as the source of the key for generating the tag for the share. Since the Security Hub can intentionally create bit strings that produce colliding tags, a key that is unknown to the Security Hub must be used.

\section{Security analysis} \label{sec:secanalysis}

We now present a security analysis of the DSKE protocol. We start by specifying a threat model. In this paper, we will assume that an adversary can perform the following actions:
\begin{itemize}
 \item access unlimited classical or quantum computing resources (information-theoretical model)\
 \item compromise a given number of the Security Hubs without the knowledge of other parties
 \item compromise given clients
 \item monitor all communication links
 \item modify all communication on all links (excluding in a denial-of-service attack)
\end{itemize}

When an adversary compromises a party, either a Security Hub or a DSKE client, we assume that he is able to access all its local data, and to have that party send messages of their choice.

In order to set the framework for the presentation and analysis of our design, we introduce in this section a number of relevant definitions. While most of those definitions are general and can refer to any key distribution protocol, we provide them specifically in the context of the DSKE system presented in this paper. Here we must distinguish between the sender (source of the secret: Alice) and the receiver(s) (the other parties to the secret: Bob, \ldots). These definitions are:
\begin{itemize}
 \item The \textit{leakage threshold} of the DSKE protocol is the minimum number of Security Hubs that must be compromised to breach confidentiality of the sender key, regardless of failure, active interference or passive monitoring on any communication paths. This may be thought of as a ``sender leakage threshold''.
 \item The \textit{injection threshold} of the DSKE protocol is the minimum number of Security Hubs that must be compromised to breach confidentiality of the receiver key, regardless of failure, active interference or passive monitoring on any communication paths. This may be thought of as a ``receiver leakage threshold''. Security Hubs of the set acceptable to the receiver are candidates for compromise.
 \item The \textit{malleability threshold} of the DSKE protocol is the minimum number of Security Hubs that must be compromised to avoid detection by any client of mismatched keys between clients, regardless of failure, active interference or passive monitoring on any communication paths. An adversary effecting matching changes in the agreed key without gaining information on the key is excluded, as this is not a vulnerability. Security Hubs in the set acceptable to the receiver are candidates for compromise.
 \item The \textit{disruption threshold} of the DSKE protocol is the minimum number of Security Hubs that must be compromised or have active interference on their communication paths before the protocol is disrupted from agreeing a key without aborting the protocol, regardless of passive monitoring of any communication paths. Security Hubs in the set selected by the sender are candidates for compromise.
\end{itemize}

\subsection{Attacks on the DSKE system} \label{subsec:attackondske}

There are several potential avenues of attack on the DSKE protocol. This section lists the most relevant ones, and the next section gives the security thresholds for the DSKE protocol in the context of the threat model described above.
\begin{itemize}
 \item \textit{Security Hub compromise}.
 This attack involves compromise of one or more Security Hubs by an adversary. Such a compromise may be total, in the sense that the adversary may have access to all data that the Security Hub has access to, and may control its operation entirely, unknown to any other parties.
 \item \textit{Client compromise}.
 This attack involves compromise of one or more clients by an adversary. Attacks on the DSKE protocol in which a compromised client is involved are self-evident, and will be considered out of the scope of this analysis.
 \item \textit{PSKM compromise}.
 This attack involves the interception and compromise of the PSKM shipped by the Security Hub to the DSKE clients. An adversary can intercept the PSKM and read the stored random data in it. A common practice for implementation of the DSKE protocol to mitigate this attack would be to require the PSKM to be tamper-proof or tamper-evident, and the data in it to be protected by symmetric encryption, with a key delivered off-line to the client on a separate channel.
 In case the PSKM compromise attack goes undetected before use by both the client and the Security Hub, this is equivalent to a full compromise of the Security Hub. In case the attack is detected by either agent before its use, it is equivalent to DoS of the Security Hub. A modification of the data on the PSKM can generate a subversion of Security Hub's client identity-verification process, which is equivalent to an undetected reading and nondelivery of the PSKM en route.
 \item \textit{Message interception without modification}.
 In this attack the adversary intercepts messages between Security Hubs and clients, or between clients, without modifying them. Since the confidential portions of the protocol are securely encrypted using a key that is known to only the communicating parties, this is only useful to an adversary who has information on the key. Thus, this attack adds no threat to the DSKE protocol.
 \item \textit{Message interception with modification}.
 In this attack the adversary intercepts messages from Security Hubs and clients or between clients, and modifies them before delivery. We note that messages to and from Security Hubs for which confidentiality is needed are encrypted using a one-time pad, with keys known only to the communicating parties, providing the necessary protection.
 Attacks on protocol parameters that are transmitted in messages would enable possibly subtle but serious security flaws. This includes modification of communicated parameters $n$ and $k$, the indices into the tables $H^A_i$ or $H^B_i$, identities, the message sender or receiver IDs, as well as other parameters such as the $x$-coordinates of Shamir's secret sharing scheme. All messages therefore must have integrity and authenticity protection. Because these parameters are included in the messages explicitly or implicitly by use of key material, they are protected against modification attacks by the use of message tags that use keys that are known only to the communicating parties. This implies that, beyond other attacks, attacks on individual messages contribute only weakness to DoS and a vanishingly low probability of subversion through tag forgery.
\end{itemize}

Other possible attacks in a similar fashion are the following:
\begin{itemize}
 \item Capturing of valid messages from one execution of a protocol and replaying them in another would allow an adversary to induce incorrect parameters. A nonce such as a key identifier that is securely bound to the protocol prevents such mixing attacks.
 \item Injection attack through Security Hubs: Compromised Security Hubs allow an adversary to modify the protocol parameters conveyed by those Security Hubs without this being detectable by $B$ through his validation of the message tag. Parameters that are duplicated for each Security Hub ($n$, $k$, key ID, final key tag) can be cross-checked with at least $k$ needing to agree, making such an attack infeasible for fewer compromised Security Hubs. If only $k$ shares are received, no inconsistency will be detected by $B$. The final key tag protects against this attack up to the injection threshold.
 \item Depletion attacks: One or more clients may strain or deplete the resources of other parties, such as Security Hub or client bandwidth, or exhaust the key material of another client. This attack can be mitigated by allowing DSKE clients to pre-set constraints on key requests from each other.
 \item Undermining trust: A potential strategy for an adversary is to discourage clients from using honest Security Hubs so that they use compromised Security Hubs. This is possible when the adversary controls a majority of the Security Hubs that are selected by $A$. More significantly, control of $A$ by the adversary allows the adversary to distribute a bad share through any Security Hub, which is indistinguishable to $B$ from that Security Hub deliberately mounting a DoS attack or malfunctioning. Such an attack by a client is not addressed in this protocol, but must be considered when formulating a strategy for excluding Security Hubs. A similar argument applies to compromise of $B$ in the adapted protocol.
\end{itemize}

In conclusion, most of the attacks on the DSKE protocol communication can be mitigated by properly encrypting and authenticating all the communications between Security Hubs and DSKE clients.

\subsection{Analysis of DSKE protocol}

The security of the DSKE protocol can be fully characterized by the four thresholds defined at the beginning of this section (leakage,  injection, malleability, disruption). As described in section \ref{subsec:attackondske}, the only real threat to the DSKE protocol comes from undetected PSKM compromise by an adversary, which is equivalent to a full Security Hub compromise. In DSKE, this risk is intrinsically mitigated by the decentralized nature of the Security Hubs. Regardless, it is clear that the first phase of the protocol, i.e., the physical shipment of the PSKM, plays a crucial role to preserve the security of the entire system.

Here we report the thresholds of the protocols together with considerations for choosing the parameters $n$ and $k$. The definitions of the security thresholds given above and the requirements for running the protocol have been chosen to make the results for the thresholds consistent across the protocols presented here. 

The leakage threshold represents the number of compromised Security Hubs that is necessary to breach confidentiality of the sender key. In the share distribution phase of DSKE, $A$ encrypts the shares that she transmits with an information-theoretically secure cipher, and each Security Hub similarly encrypts its share for transmission to $B$. By the definition of the secret sharing scheme, at least $k$ of the Security Hubs that $A$ chooses must be compromised before any information of the key that $A$ takes as agreed is gained by an adversary. Therefore, the leakage threshold is $k$.

The injection threshold represents the number of compromised Security Hubs that is necessary to breach confidentiality of the receiver key. $B$ decrypts the shares that he receives from the Security Hubs with an information-theoretically secure cipher, and each Security Hub decrypts its share for transmission to $B$ similarly, which results in attacks on the communication link providing no information to an adversary. By the definition of the secret sharing scheme, at least $k_B$ of the Security Hubs that $B$ will accept for the protocol must be compromised before any information of the key that $B$ takes as agreed is gained by an adversary, where $k_B$ is the minimum that $B$ will accept for the protocol.

The malleability threshold represents the number of compromised Security Hubs that is necessary to have Alice and Bob mistakenly agree on two different keys. $B$ validates the shares that he receives with an information-theoretically secure message tag, and each Security Hub validates its share for transmission to $B$ similarly, which results in attacks on the communication link providing no information to an adversary. Hence, at least $k_B$ of the Security Hubs that $B$ will accept for the protocol must be compromised before any key that $B$ takes as agreed can be modified by an adversary, where $k_B$ is the minimum that $B$ will accept for the protocol.

Finally, the disruption threshold represents the number of compromised or disrupted Security Hubs needed to force Alice and Bob to abort the protocol without having achieved agreement on a key. For the protocol to be disrupted, one of the following conditions must hold:
\begin{enumerate}
 \item The number of Security Hubs that successfully deliver their shares to $B$ must be less than $k$. For this, no fewer than $n - k + 1$ of these Security Hubs or their communication links must be disrupted or compromised, where $n$ and $k$ are the values selected by $A$.
 \item There must be at least one internally consistent group of Security Hubs, large enough to be accepted by $B$, that produces a different key and hence, with high probability, also a different key tag. The transmission of the key tag with the key instruction message prevents the adversary from using a group that overlaps with the genuine group. The key tag is transmitted via an independent path for the adapted protocol, but the vanishing probability of forgery allows $B$ to use it to find the valid share combinations. This would need at least $k_B$ of the Security Hubs that $B$ will accept to be compromised.
\end{enumerate}

This leads to the disruption threshold being reached if either $n - k + 1$ of the Security Hubs are compromised or disrupted, or $k_B$ of the Security Hubs that $B$ accepts being compromised by the adversary. So this threshold is the lesser of $n - k + 1$ and $k_B$.

We can summarize the thresholds in the following bullets:
\begin{itemize}
 \item Leakage threshold: $k$ (value chosen by $A$)
 \item Malleability threshold: $k_B$ (minimum value of $k$ that $B$ will accept)
 \item Injection threshold: $k_B$ (minimum value of $k$ that $B$ will accept)
 \item Disruption threshold: the lesser of $n - k + 1$ (values chosen by $A$) and $k_B$.
\end{itemize}

The sum of the leakage and disruption thresholds for the DSKE protocol is $n + 1$ (provided only that $k_B \ge n - k + 1$), independently of the secret sharing threshold $k$. That this cannot be improved is evident from the definition of an $(n,k)$ secret sharing scheme: leakage cannot be prevented with the compromise of $k$ or more shares, and reconstruction is impossible (implying disruption) with the omission of $n - k + 1$ or more shares.

\subsubsection{Choice and tradeoff of security parameters}

The tradeoff between the disruption threshold and the other thresholds is clear for a given $n$: as the disruption threshold (and hence robustness) increases, the other thresholds decrease. Where DoS is a concern, the value $k$ must be reduced below $n$ at the sacrifice of the other security margins. Where other security criteria (such as confidentiality and authenticity) are paramount, $k$ should be made large. The minimum $k$ that $B$ will accept (i.e., $k_B$) cannot be made too low though, since the gain in robustness against disruption through active attack through compromised Security Hubs reverses near this extreme. 

For a given $k$, as $n$ increases, there is more opportunity for compromise due to the greater attack surface. In addition, as the pool of accepted Security Hubs increases with both $k$ and $n$ held constant, there is similarly a reduction in security, primarily in the form of getting $B$ to accept an incorrect key, including one known to the adversary. All security metrics may be improved by increasing the parameters $n$ and $k$ in a coordinated way, assuming that the quality (robustness against compromise and failure) of the Security Hubs is retained. Other clients in the network also pose some threat, largely inasmuch as that they may themselves be compromised, thus posing a risk of an attack on the credibility of Security Hubs. This type of attack is not discussed in this paper.

\section{DSKE implementation} \label{sec:implementation}

\subsection{Share performance} \label{subsec:shareimpl}

We constructed a prototype implementation of our system based on the General DSKE Protocol. As described in Sec. \ref{sec:dskeprot}, we used Shamir's secret sharing scheme over $\mathrm{GF}(2^8)$, with two clients and multiple Security Hubs communicating over TCP/IP with only the unidirectional messages described. This implementation did not include the adapted protocol described above, but it allowed us to run performance tests on the system.

In particular, we demonstrate performance for a range of $n$ and $k$ ($1 \le n \le 20$, $1 \le k \le n$) for the number of shares $s$ in the range $k \le s \le n$ received by Bob, where zero or more of the $n$ shares are corrupted, either with a mismatched tag (as for data corruption in transit) or with a matching tag (as for a compromised Security Hub attacking the system). The agreed secret length was set to 8 megabits.

\begin{figure}[ht]
\includegraphics[width=1\columnwidth]{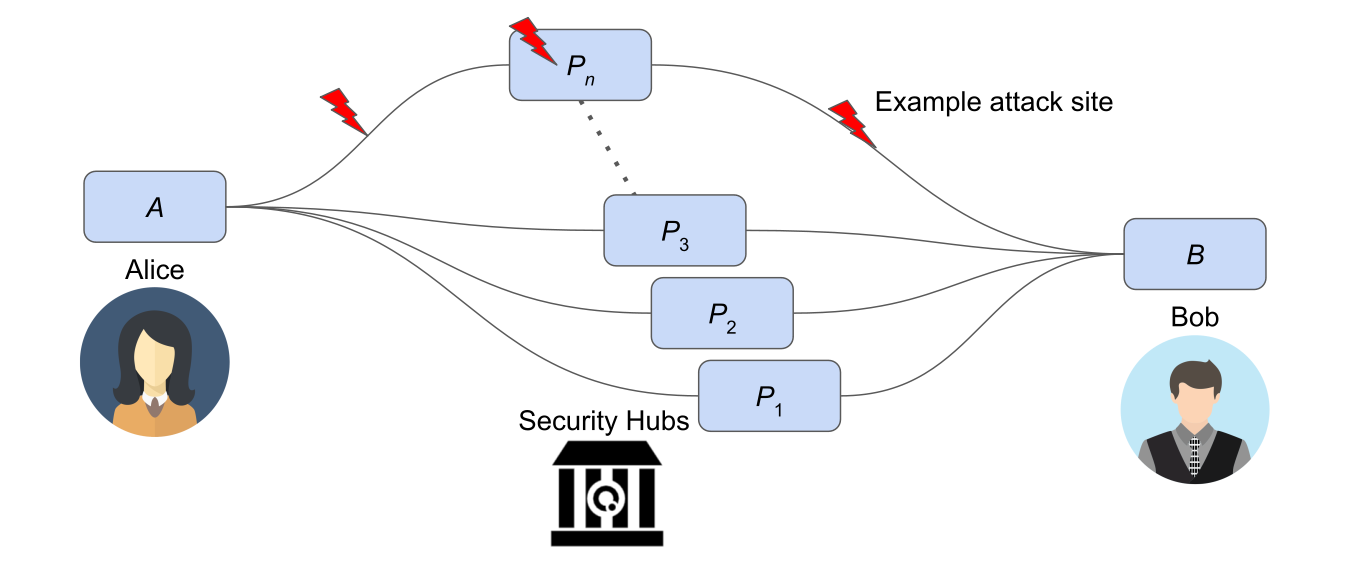}
\caption{\textit{Communication paths in the DSKE protocol after PSKM distribution. Only one message propagates on each arrow. Examples of attack sites modeled with the implementation are shown.}}
\label{fig:attack}
\end{figure}

This implementation validates that the scheme functions throughout parameter space for small $n$ and that the disruption threshold is sharp for small $n$ when the Security Hubs do not exploit knowledge of the key: whenever the disruption threshold is not reached, reconstruction arrives at the correct key. It does not provide a complete validation of the security properties, which requires a more rigorous approach using proof.

\begin{figure}[ht]
\includegraphics[width=1\columnwidth]{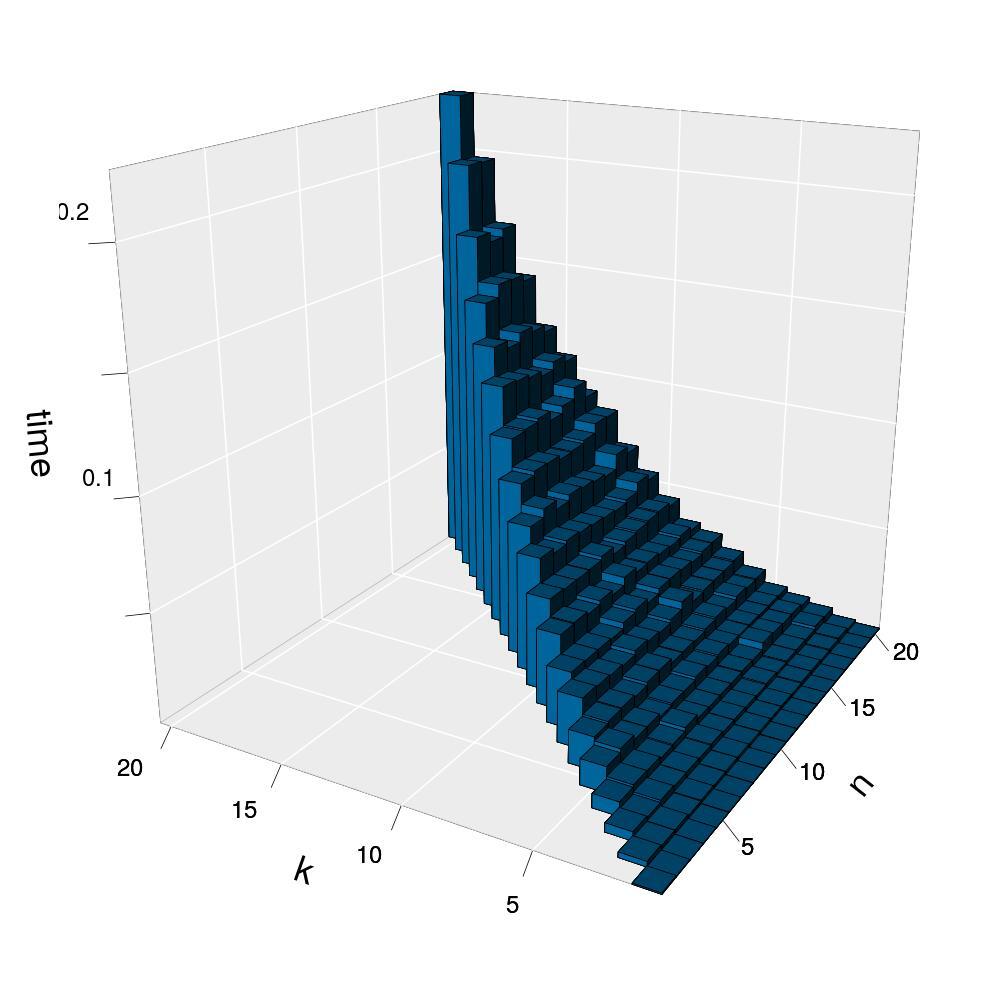}
\caption{\textit{Scaling behavior of processing rate of 8\nobreakspace Mbit keys agreed as a function of $n$ and $k$.  Time is in seconds.}}
\label{fig:scalen}
\end{figure}

Processing cost for Alice and Bob as affected by scaling of $n$ and $k$ is shown in Fig. \ref{fig:scalen}, with their processing rates being nearly indistinguishable. We implemented the receiver validation for quickest operation when the system is not under active attack, namely by reconstructing the secret from only $k$ shares and then validating the result against the key tag, only proceeding to another combination of shares if this fails. No attempt was made at full optimization.

We omitted communication and data retrieval costs, which can be separately estimated and will scale approximately linearly with the amount of data communicated by the protocol, including the message cryptographic primitives. The results here inherently reflect design choices that we made. Here, secret reconstruction included first deriving the polynomial coefficients, and then using these to derive the secret and the other shares as needed. The step of deriving the shares from the coefficients was found to take nearly negligible time compared to the generation of the coefficients. These results demonstrate an implementation where the processing cost at each of $A$ and $B$ scales close to proportionally with $k^2$ with minimal dependence on $n$; this is sufficient to show that implementations for realistic values of $n$ and $k$ can be performant. Example parameter values might be $n = 9$ and $k = 5$, with share processing time in the order of 1\nobreakspace ms/Mbit at each client for our implementation.

\subsection{DSKE integration in real networks} \label{subsec:realnets}

The DSKE implementation described in section \ref{subsec:shareimpl} is very useful to understand the performance of the system for different values of $n$ and $k$, without taking into account details related to routing time, network traffic, and other sources of complexity that arise in a real network environment. In this section instead, we integrate the Simple DSKE Protocol for protecting real traffic in a network.

A DSKE client can be integrated in different layers of the OSI model for network architectures. In this specific example, we use DSKE to provide key material for protecting a layer 3 Virtual Private Network (VPN) tunnel. As VPN we use the WireGuard\textsuperscript{\textregistered} network tunnel\footnote{"WireGuard" is a registered trademark of Jason A. Donenfeld.} \cite{donenfeld2017wireguard}. WireGuard is a kernel virtual network interface for Linux, which uses ChaCha20-Poly1305 authenticated-encryption \cite{nir2018chacha20} to encapsulate packets of protocols such as TCP, transported in UDP packets. The keys used for the encapsulation are exchanged by WireGuard peers via the Curve25519 ECDH function \cite{bernstein2006curve25519}. To our knowledge, the best known attack on ChaCha20 uses Grover’s algorithm running on a quantum computer, which is mitigated by a doubling of the key length \cite{grover1996fast}. The Curve25519 function is vulnerable to Shor's algorithm running on a quantum computer \cite{langley2016elliptic}.

In order to mitigate this threat, WireGuard supports an additional pre-shared key amongst the communicating parties. This adds a layer of symmetric cryptography for deriving the session key. We use DSKE to provide this pre-shared key, refreshed every two minutes to match the WireGuard handshake interval \cite{donenfeld2017wireguard}. Rapidly refreshing the session key has been established as increasing security, including by limiting the amount of data protected by any single key, by presenting the attacker with a dynamic target, and by allowing systems to treat the key as ephemeral (erasing any record of it immediately after use). The upper practical limit on the rate at which the preshared key for WireGuard may be refreshed is set by its handshake interval.

To demonstrate the performance of DSKE, we implemented an OTP mini-VPN module that provides one-time-pad encryption and authentication using DSKE as a key source. This module is implemented as a Linux TUN device that captures and manages internet packets from a designated network interface and processes these. It encrypts and decrypts packets depending on whether they are outbound or inbound by using a Vernam cipher and authenticates and validates these using Poly1305. It also manages synchronization of the keystream. An encrypted packet is transported using UDP, with the plaintext packet being from any ethernet protocol. In the simple form used in this implementation, it does not provide reliability or resistance to DoS attacks.

\begin{figure}[ht]
\includegraphics[width=1\columnwidth]{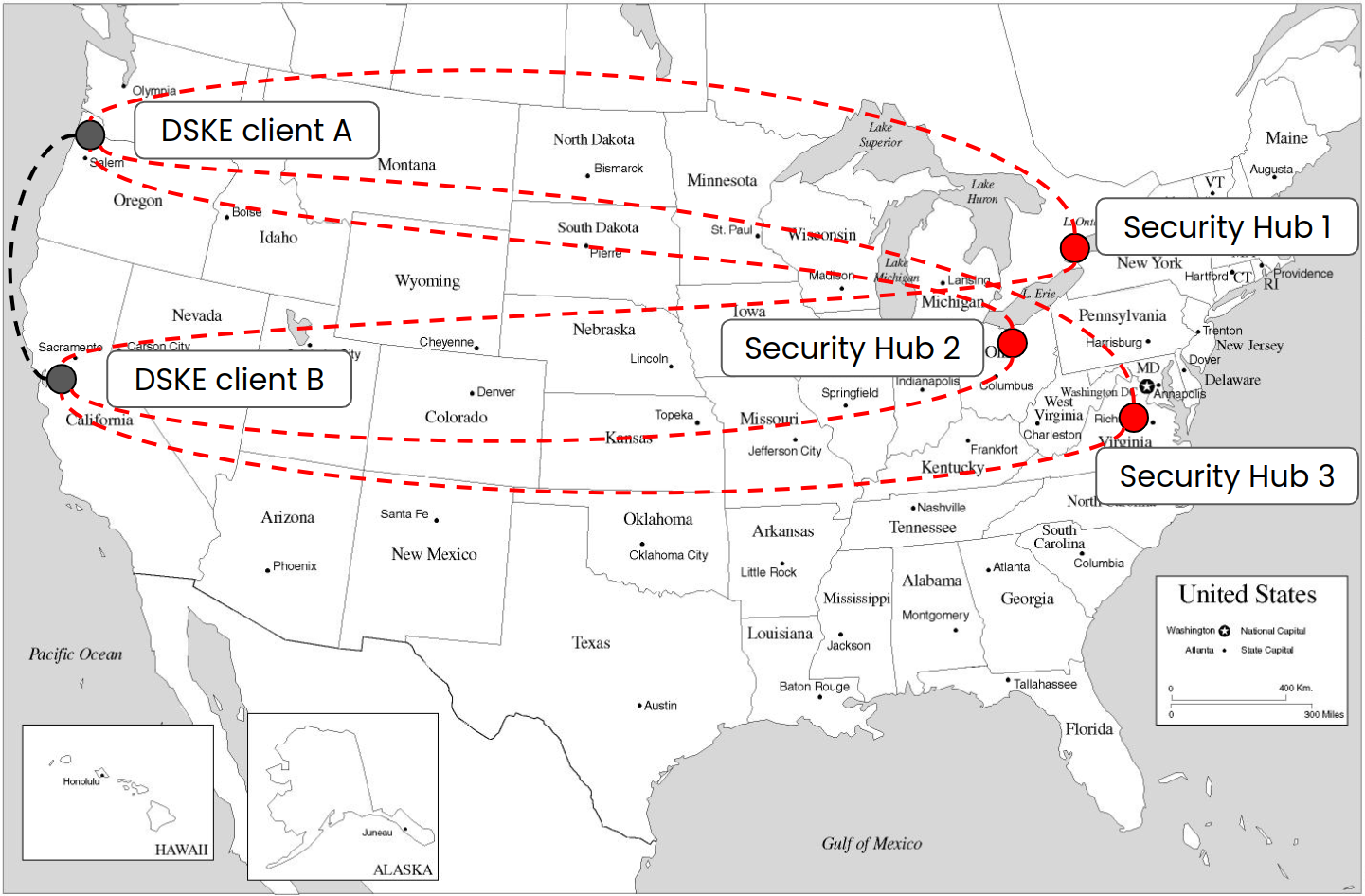}
\caption{\textit{DSKE network topology as implemented, showing node locations.}}
\label{fig:map}
\end{figure}

We deployed the above system on Amazon Web Server (AWS) nodes using the Simple DSKE Protocol with three Security Hubs and two clients, Alice and Bob. The physical key delivery in this case was replaced by a simple data transfer, as our objective was only to measure performance. The five AWS nodes were located as in Fig. \ref{fig:map}. While in this section we focus on the case $n = k = 3$, one can refer to the previous subsection to get estimates on how different choices of $n$ and $k$ would impact the performance of the DSKE key generation process.

We measure throughput and latency between Alice and Bob for the four following channel configurations:
\begin{itemize}
 \item C1: without WireGuard (no protection)
 \item C2: with WireGuard
 \item C3: with WireGuard with keys from DSKE every 2 minutes
 \item C4: with the OTP mini-VPN as a Linux TUN device.
\end{itemize}

Throughput is measured with the \textit{iperf} Linux tool, while latency is measured using the \textit{ping} command. For each of the configurations, we repeat the measurement ten times and report the average and standard deviation in Table \ref{table:perf} to give an indication of the impacts.

\begin{table}[h]
\begin{tabular}{ccc}
\textbf{Channel} & \textbf{Throughput [Mbit/s]} & \textbf{Latency [ms]} \\
C1    & $ 646 \pm  18 $    & $ 19.4 \pm 0.1 $ \\
C2    & $ 733 \pm  77 $    & $ 20.8 \pm 0.1 $ \\
C3    & $ 779 \pm 110 $    & $ 20.2 \pm 0.1 $ \\
C4    & $  53 \pm 1.6 $    & $ 63.8 \pm 0.9 $
\end{tabular}
\break\break
\caption{\textit{Mean and standard deviation of the throughput and latency for the channels C1--C4.}} \label{table:perf}
\end{table}

Channels C1--C3 show inconclusive differences between cases, which is consistent with the low rate of keys required by WireGuard from DSKE being expected to have minimal impact. Channel C4 shows that DSKE is able to supply keys with $n = 3$ Security Hubs at a continuous rate of at least 50\nobreakspace Mbit/s. This would be sufficient for approximately 20 video calls in parallel. 1\nobreakspace TB PSKM data from each Security Hub would be enough for 48 hours of standard definition video call, 2 years of VoIP, or 10 million emails of average size. The Linux TUN interface may be a significant factor limiting the data rate, which would be mitigated by integrating the OTP mini-VPN into the Linux kernel.

\section{Conclusions} \label{sec:conclusion}

We focused on the problem of distributing cryptographic keys among clients with a system that achieves information-theoretic security and robustness. Our proposal, which we call Quantum Key Infrastructure (DSKE), relies on a number of service providers, called Security Hubs. Security Hubs do not need to be trusted individually by DSKE clients, and the DSKE protocol is able to tolerate a certain fraction of compromised Security Hubs, decided by the clients. DSKE achieves information-theoretic security at the logistic cost of physically delivering Pre-Shared Key Modules (PSKMs) between Security Hubs and DSKE clients. The system is fully scalable and can be implemented with limited computational resources. The DSKE system can be easily integrated in the Internet stack, and finds important use cases in security of networks, end-user devices (such as mobile phones), and embedded devices. For large embedded device systems, for example networks of distributed sensors or IoT networks, DSKE offers the unique advantage of being computationally light and scalable to very large systems. A proof-of-principle demonstration of a DSKE-enabled VPN is also performed, demonstrating its high throughout and low latency.



\begin{acks}
The research reported in this paper was supported by the Connaught Innovation Award, the Borealis AI Graduate Fellowship, and Mitacs Accelerate.

We are grateful for the assistance in preparing the paper given by Andrew Csinger, Damian Bozewicz, and Jie Lin.
\end{acks}

\bibliographystyle{ACM-Reference-Format}
\bibliography{DSKE}

\end{document}